\documentclass{PoS}
\newcommand{\Det}{{\rm Det}}
\newcommand{\Tr}{{\rm Tr}}
\newcommand{\tr}{{\rm tr}}

\newcommand{\be}{\begin{equation}}
\newcommand{\ee}{\end{equation}}
\newcommand{\bea}{\begin{eqnarray}}
\newcommand{\eea}{\end{eqnarray}}
\newcommand{\ba}{\begin{array}{l}}
\newcommand{\ea}{\end{array}}

\newcommand{\re}[1]{(\ref{#1})}

\usepackage[normalem]{ulem} 
\renewcommand\sout{\bgroup \color[rgb]{1,0,0} \ULdepth=-.5ex \ULset}

\title{Heavy-light quarks interactions in QCD vacuum }

\ShortTitle{Heavy-light quarks interactions in QCD vacuum }

\author{\speaker{Mirzayusuf Musakhanov}%
        \\
        National University of Uzbekistan\\
        E-mail: \email{musakhanov@gmail.com}}

\abstract{
QCD vacuum instantons induce very strong interactions between light quarks, which generate large dynamical light quark mass M for initially almost massless quarks and can bound these quarks to produce almost massless pions in accordance with the spontaneous breaking of chiral symmetry ($S\chi$SB). On the other hand, the QCD vacuum instantons generate heavy-light quark interactions terms, which are responsible for the effects of $S\chi$SB in a heavy-light quark system. Summing the re-scattering series that lead to the total light quark propagator and making few further steps, we get the fermionized representation of low-frequencies light quark determinant in the presence of the quark sources, which is relevant for our problems. The next important step in the line of this strategy is to derive the equation and calculate the heavy quark propagator in the instanton media and in the presence of light quarks. This one provide finally the heavy and $N_f$ light quarks interaction term. As an example, we derive heavy quark-light mesons interaction term for the $N_f=2$ case. If we take the average instanton size $\rho=$0.35 fm, and average inter-instanton distance R=0.856 fm from our previous estimates, we obtain at LO on $1/N_c$ expansion dynamical light quark mass M = 570 MeV and instanton media contribution to heavy quark mass $\Delta M$=148 MeV. These factors define the coupling between heavy and light quarks and, certainly, between heavy quarks and light mesons. We will apply this approach to heavy quark and heavy-light quark systems. 
}

\FullConference{XXII International Baldin Seminar on High Energy Physics Problems \\
                 15-20 September, 2014\\
                 JINR, Dubna, Russia}

\begin{document}

\section{Introduction}
One of the most prominent advances of the QCD instanton vacuum model 
is the correct description of the spontaneous breaking of chiral
symmetry ($S\chi$SB), which is responsible for properties of most
light hadrons and nuclei. Instantons induce
very strong interactions between light quarks, which generate
the dynamical quark mass $\sim 400\,\,\mathrm{MeV}$ 
 for initially almost massless quarks and can bound
these quarks to produce almost massless pions in accordance with
$S\chi$SB.   In the instanton picture $S\chi$SB is due to the
delocalization of single-instanton quark zero modes in the instanton medium. 

The instanton vacuum field is assumed as a
superposition of $N_{+}$ instantons and $N_{-}$ antiinstantons: 
\begin{eqnarray}
A_{\mu}(x)=\sum_{+}^{N_{+}}A_{\mu}^{I}(\zeta_{+},x)+\sum_{-}^{N_{-}}A_{\mu}^{A}(\zeta_{-},x). 
\label{A}
\end{eqnarray}
Here, $\zeta=(\rho,z,U)$ denote respectively
(anti)instanton collective coordinates-- 
size, position and color orientation (see reviews
~\cite{Diakonov:2002fq,Schafer:1996wv}). One of the advantages of
the  instanton vacuum model is that it is characterized by only two parameters:
the average instanton size $\rho$ and the average
inter-instanton distance $R$. The estimates of these
quantities are 
\begin{eqnarray}
 &  & \rho\simeq0.33\, \mathrm{fm},\, R\simeq1\,
      \mathrm{fm},
      \mbox{(phenomenological)}~~
      \mbox{\cite{Diakonov:2002fq,Schafer:1996wv}}, 
\label{classicalParameters}\\    
 &  & \rho\simeq0.35\, \mathrm{fm},\, R\simeq0.95\,
      \mathrm{fm},\mbox{(variational)}~~\mbox{\cite{Diakonov:2002fq}},
\nonumber   \\ 
 &  & \rho\simeq0.36\, \mathrm{fm},\, 
 R\simeq0.89\, \mathrm{fm},~\mbox{(lattice)}~\mbox{\cite{lattice}}.
 \nonumber \end{eqnarray}

Our estimates~\cite{Goeke:2007bj} (in the  instanton vacuum model with
 $1/N_c$ corrections taken into account) from
the pion decay constant
$F_{\pi,{m=0}}=88\,\mathrm{ MeV}$ and the quark condensate
$\langle\bar qq\rangle_{m=0}=-(255
\,\mathrm{MeV})^3$ in the chiral limit~\cite{Leutwyler:2001hn}
are given as:   
\bea
R\approx 0.856\,\mathrm{fm},\,\,\,  \rho\approx 0.35\,\mathrm{fm}.
\label{rhoR}
\eea 
A recent computer
simulation~\cite{Cristoforetti:2006ar} of a
current mass dependence of QCD observables within the instanton
liquid model show that the best correspondence to the lattice
QCD data is obtained  as $R\approx
0.76\,\mathrm{fm},\,\,\,  \rho\approx 0.32\,\mathrm{fm}$. 
Thus within 10--15$\%$ uncertainty, different approaches 
give similar estimates. In the following we will use our values for
$R,\rho$~\re{rhoR}. 

While the instantons are absolutely important for the light quark physics, for 
 the heavy quarks even the charmed quark mass $m_{c}\sim1.5$~GeV is
 essentially larger than the typical parameters of the instanton
 media--the inverse instanton size $\rho^{-1}\approx600$~MeV and the
 inter-instanton distance $R^{-1}\approx200$~MeV and thus the
 quark mass mainly determines the dynamics of the heavy
 quarks~\cite{Diakonov:1989un,Chernyshev:1995gj}. 
On the other hand, the QCD vacuum instantons generate heavy-light quark interactions terms 
which are responsible for the effects of $S\chi$SB in a
heavy-light quark system. Our aim is to derive these
interaction terms. First, we have to take into account light quarks
effects in the partition function. 
\section{Light quark determinant with the quark sources term}
We start from the splitting of
 the total quark determinant to the low and high frequencies
parts as $\Det=\Det_{{\rm high}}\cdot\Det_{{\rm low}}$, where $\Det_{{\rm high}}$
gets a contribution from fermion modes with Dirac eigenvalues from
the interval $M_{1}$ to the Pauli--Villars mass $M$, and $\Det_{{\rm low}}$
is accounted eigenvalues less than $M_{1}$. The product of these
determinants is independent  of the scale
$M_{1}$. However,
we may calculate 
both of them only approximately. There is a week dependence of the
product on $M_{1}$in the wide range of $M_{1}$, which serves as
a check of the approximations ~\cite{Diakonov:1995qy}.

The high-momentum part ${\Det}_{{\rm high}}$ can be written as a
product of the determinants in the field of individual instantons,
while the low-momentum one ${\Det}_{{\rm low}}$ has to be treated
approximately, would-be zero modes being taken into account only.

The next step is to compute the light
quark propagator in the instanton media. Our main assumption is the
interpolation formula~\cite{Goeke:2007bj,Musakhanov}: 
\begin{eqnarray}\label{Si}
S_{i}=S_{0}+S_{0}\hat{p}\frac{|\Phi_{0i}><\Phi_{0i}|}{c_{i}}\hat{p}S_{0},\,\,\, S_{0}=\frac{1}{\hat{p}+im},\,\,\,
c_{i}=im<\Phi_{0i}|\hat{p}S_{0}|\Phi_{0i}> .
\end{eqnarray}
 The advantage of this interpolation is shown by the projection of
$S_{i}$ to the zero-modes: 
\begin{eqnarray}
S_{i}|\Phi_{0i}>=\frac{1}{im}|\Phi_{0i}>,\,\,\,<\Phi_{0i}|S_{i}=<\Phi_{0i}|\frac{1}{im}
\end{eqnarray}
as it must be, while the similar projection of $S_{i}$ given by
Ref.~\cite{Diakonov:1995qy} has a wrong component, negligible only in 
the $m\rightarrow0$ limit.

Summing the re-scattering series that
lead to the total quark propagator and making few further
steps, we get the fermionized representation of low-frequencies light
quark determinant in the presence of the quark sources, which is
relevant for our problems, in the form~\cite{Goeke:2007bj,Musakhanov}: 
\bea
\label{part-func}
&&{\Det}_{\rm low}\exp(-\eta^{+}S\eta)=
\\\nonumber
&&\int\prod_{f}D\psi_{f}D\psi_{f}^{\dagger}\exp\int\sum_{f}\left(\psi_{f}^{\dagger}(\hat{p}\,+\, im_{f})\psi_{f}+\psi_{f}^{\dagger}\eta_{f}+\eta_{f}^{+}\psi_{f}\right)
\prod_{f} \prod_{\pm}^{N_{\pm}}V_{\pm,f}[\psi^{\dagger},\psi],
\end{eqnarray}
 where 
 \begin{eqnarray}
V_{\pm,f}[\psi^{\dagger},\psi] = i\int
   d^{4}x\left(\psi_{f}^{\dagger}(x)\,
   \hat{p}\Phi_{\pm,0}(x;\zeta_{\pm})\right)\int
   d^{4}y\left(\Phi_{\pm,0}^{\dagger}(y;\zeta_{\pm})(\hat{p}\,\psi_{f}(y)\right). 
\label{V}\end{eqnarray}
The averaging over collective coordinates $\zeta_{i,\pm}$ of
${\Det}_{\rm low}\exp(-\eta^{+}S\eta)$  is a rather simple procedure, 
since the low density of the instanton medium
($\pi^{2}\left(\frac{\rho}{R}\right)^{4}\sim0.1$) allows us to average
over positions and orientations of the instantons independently. This
one leads to the light quark partition function $Z[\eta,\eta^+]$. From
\re{part-func} at $N_f=1$ and $N_\pm=N/2$ it is exactly given by  
\bea
&&Z[\eta,\eta^+]=e^{-\eta^+\left(\hat p \,+\, i(m+M(p))\right)^{-1}\eta}
\exp\left[\Tr\ln\left(\hat p \,+\, i(m+M(p))\right)+N\ln\frac{N/2}{\lambda}-N\right],
\label{Z}
\\
&&N=\Tr\frac{iM(p)}{\hat p \,+\, i(m+M(p))},\,\,\, M(p)=\frac{\lambda}{N_c}(2\pi\rho F(p))^2.
\label{M}
\eea
Here the form-factor $$F(p) 
= 2z \left(I_0 (z) K_1 (z)- I_1 (z) K_0 (z)
-\frac{1}{z} I_1 (z) K_1 (z)\right)$$
($I_0$, $I_1$, $K_0$, $K_1$ are the modified Bessel
functions, $z=p\rho/2$) 
is given by Fourier-transform of the
zero-mode. The coupling  $\lambda$ and the dynamical quark mass $M(p)$
are defined by  Eq. \re{M}. 

 At $N_f >1$ and in the saddle-point approximation (no meson loops
 contribution) $Z[\eta_f,\eta_f^+]$ has a similar form to 
 Eq. \re{Z}.   
 \section{Heavy quark propagator} 
The next important step in the line of this strategy is to calculate the heavy quark propagator
in the instanton media and in the presence of light quarks. We will
extend the equation for the heavy quark propagator in the instanton
media previously derived in
Refs.~\cite{Diakonov:1989un,Pobylitsa:1989uq}.   
 
The heavy quark Lagrangian in the external gluon field and in Euclid
space is given by 
$
L_H=\Psi^+(\hat P+im_H)\Psi,\,\,\,P=p-gA.
$
We make a Foldy-Wouthuysen transformation
accordingly\cite{Chernyshev:1995gj}: 
$
\Psi(x)=\exp(-m_H\gamma_4 x_4+O(1/m_H))Q(x),
$
which leads to 
\bea
L_H=\Psi^+(\hat P+im_H)\Psi=Q^+\gamma_4 P_4 Q +Q^+Q_1 Q,\,\,\, 
Q_1=\frac{{\vec P}^2}{2m_H}-\frac{\vec\sigma\vec B}{2m_H},\,\,\, \vec
B=rot \vec A .  
\eea
In the present case we neglect  $O(1/m_H)$ terms and
define the heavy quark propagator as: 
\bea
&&S_H=\frac{1}{Z}\int \prod_{f}D\psi_{f}D\psi_{f}^{\dagger}\exp\int\sum_{f}\left(\psi_{f}^{\dagger}(\hat{p}\,+\, im_{f})\psi_{f}\right)
 \prod_{\pm}^{N_{\pm}}<{\prod_{f}V_{\pm,f}[\psi^{\dagger},\psi]}> w[\psi,\psi^\dagger],
\\ \nonumber
&&<{\prod_{f}V_{\pm,f}[\psi^{\dagger},\psi]}>\equiv\int d\zeta_\pm\prod_{f}V_{\pm,f}[\psi^{\dagger} ,\psi ],
\\ \nonumber
&&w[\psi,\psi^\dagger]
=\left\{\prod_{\pm}^{N_{\pm}}<\prod_{f}V_{\pm,f}[\psi^{\dagger},\psi]>\right\}^{-1}\int\prod_{\pm}^{N_{\pm}} d\zeta_\pm
\left\{\prod_{\pm}^{N_{\pm}}V_{\pm,f}[\psi^{\dagger} ,\psi ]\right\}\frac{1}{\theta^{-1}-\sum_i a_i},
\\\nonumber
&&<t|\theta|t'>=\theta(t-t'),  <t|\theta^{-1}|t'>=-\frac{d}{dt}\delta(t-t'),
a_i(t)=iA_{i,\mu}(x(t))\frac{d}{dt}x_\mu(t) .
\eea
 Accordingly, we derive the inverse of $w[\psi^\dagger,\psi]$
   as~\cite{Diakonov:1989un}  
\bea
&&w^{-1}[\psi,\psi^\dagger]=\theta^{-1} + \frac{N}{2}\sum_\pm \frac{1}{<{\prod_{f}V_{\pm,f}[\psi^{\dagger},\psi]}>}
\int d\zeta_\pm \prod_{f}V_{\pm,f}[\psi^{\dagger},\psi]\left( \theta-a_\pm^{-1}\right)^{-1}+ O(N^2/V^2)
\nonumber\\
&&=\theta^{-1} - \frac{N}{2}\sum_\pm \frac{1}{<{\prod_{f}V_{\pm,f}[\psi^{\dagger},\psi]}>}\Delta_{H,\pm}[\psi^{\dagger},\psi ] + O(N^2/V^2),
\eea
where 
\bea
\Delta_{H,\pm}[\psi^{\dagger},\psi ]=\int d\zeta_\pm\prod_{f}V_{\pm,f}[\psi^{\dagger},\psi]\theta^{-1}(w_\pm-\theta)\theta^{-1}
\eea
represent the interactions of heavy and $N_f$ light quarks and $
w_\pm=\frac{1}{\theta^{-1}-a_\pm}$ - heavy quark propagator in the
single (anti)instanton field.  Finally we get the quark propagator in
the instanton media with account of light quarks as 
\bea
\label{SH1}
 S_H=\frac{1}{\theta^{-1} - \lambda\sum_\pm
   \Delta_{H,\pm}[\frac{\delta}{\delta\eta}
   ,\frac{\delta}{\delta\eta^+}] } 
 \exp\left[-\eta^+\left(\hat p \,+\, i(m+M(p))
   \right)^{-1}\eta\right]_{|_{\eta=\eta^+=0}} .
\eea 
We approximate  Eq.\re{SH1}, neglecting by overlapping
diagrams as  
\bea 
&&S_H^{-1}\approx \theta^{-1} - \lambda\sum_\pm\Delta_{H,\pm}[\frac{\delta}{\delta\eta} ,\frac{\delta}{\delta\eta^+}] 
 \exp\left[\eta^+\left(\hat p \,+\, i(m+M(p))\right)^{-1}\eta\right]_{|_{\eta=\eta^+=0}}
\label{SH2}
\\
&&= \theta^{-1}  
- i\tr\int \frac{d^4 k_1}{(2\pi)^4} \frac{\lambda(2\pi\rho )^2 F^2(k_1 ) }{N_c(\hat k_1 \,+\, i(m+M(k_1)))} 
 \frac{1}{2N_c}\sum_\pm\int d^4z_\pm  \tr_c\left(\theta^{-1}(w_\pm-\theta)\theta^{-1}\right)
 \nonumber\\
 &&=\theta^{-1} - \frac{N}{2VN_c}\sum_\pm\int d^4z_\pm  \tr_c\left(\theta^{-1}(w_\pm-\theta)\theta^{-1}\right).
\label{SH3}
\eea
Then, in this approximation  Eq. \re{SH3} exactly coincides
with the similar one from~\cite{Diakonov:1989un}. 

\section{Light-heavy quarks interaction term}
Now rewrite  Eq. \re{SH1} by introducing heavy quark fields
$Q,Q^\dagger$: 
\bea
&& S_H=e^{\left[-\Tr\ln\left(\hat p \,+\, i(m+M(p))\right)\right]}\int D\psi D\psi^{\dagger}  D Q D Q^\dagger \,\,Q \, Q^\dagger\,\,\exp\left[\left(\psi^{\dagger}(\hat p +i(m+M(p)))\psi\right)\right. 
\\\nonumber
&&\left.+  Q^\dagger\left(\theta^{-1} - \lambda\sum_\pm\Delta_{H,\pm}[\psi^{\dagger},\psi ] \right)Q-\Tr\ln\left(\theta^{-1} - \lambda\sum_\pm\Delta_{H,\pm}[\psi^{\dagger},\psi ]\right)\right],
\eea
where the third term represent the (negligible) contribution of the heavy
quark loops, while the second one is the heavy and light quarks interaction action.

The heavy and $N_f$ light quark interaction term explicitly
is given by the
expression: 
\begin{eqnarray}
&&S_{Qq}= - \lambda\sum_\pm Q^\dagger\Delta_{H,\pm}[\psi^{\dagger},\psi ]Q=  
- i\lambda\sum_\pm\int d^4z_\pm dU_\pm \prod_{f=1}^{N_f}\frac{d^4 k_f}{(2\pi)^4}\frac{d^4 q_f}{(2\pi)^4}  \exp(i(q_f-k_f)z_\pm)
\nonumber\\
&&\times \frac{(2\pi\rho )^2 F(k_f )F(q_f )}{8}\psi_{f,a_f \alpha_f}^+(k_f)
(\gamma_{\mu_f}\gamma_{\nu_f} \frac{1\pm\gamma_5}{2})_{\alpha_f \beta_f}
(U^{a_f}_{\pm,i_f}(\tau^{\mp}_{\mu_f}\tau^{\pm}_{\nu_f})^{i_f}_{j_f} U^{\dagger j_f}_{\pm,b_f} \psi^{b_f}_{f,\beta_f}(q_f)
\nonumber\\
\label{SQq}
&&\times Q_{a_3}^+ U^{a_3}_{\pm,i_3}\left(\theta^{-1}(w_\pm-\theta)\theta^{-1}\right)_{j_3}^{i_3}  U^{\dagger j_3}_{\pm,b_3}  Q^{b_3} .
\end{eqnarray}
It is evident that the integration over $z$ leads to the
energy-momentum conservation delta-function, while the integration
over color orientation provides the specific structure of the
interaction terms. Also, each light quark leg is accompanied by the
form-factor $F=F(k\rho)$, which is localized at the region $k\rho\le
1,$ as expected. 
 
{\bf Details of $Q^+\left[\theta^{-1}(w_\pm-\theta)\theta^{-1}\right]Q$.}

We take initial and final positions of a heavy quarks at  $x=(\vec
x,t_1)$ and $x'=(\vec x,t_2)$ 
and have
\begin{eqnarray}
<\vec x,t_2|\left[\theta^{-1}(w_\pm-\theta)\theta^{-1}\right]|\vec x,t_1> =\frac{d}{dt_1}\frac{d}{dt_2}\Theta(t_2-t_1)\left[P\exp\left( i\int_{t_1}^{t_2}A_{\pm,4}dx_4\right)-1\right] .
\end{eqnarray}
The next step is to calculate P-exponent taking singular gauge for the
(anti)instanton at the position $z$. 

Then, representing $Q(\vec x,t)=\int \frac{d\omega d^3\vec p}{(2\pi)^4}Q(\vec p,\omega)\exp[i(\vec p\vec x+\omega t)]$ we find
\bea
&&\int dt_1dt_2d^3x Q^+(\vec x,t_2)<\vec x,t_2|\left[\theta^{-1}(w_\pm-\theta)\theta^{-1}\right]|\vec x,t_1>Q(\vec x,t_1)
\\
\nonumber
&&=
\int\frac{d\omega_1 d^3 p_1}{(2\pi)^4}\frac{d\omega_2 d^3 p_2}{(2\pi)^4} e^{i\vec p\vec z}
Q^+(\vec p_2,\omega_2)
  (J_0(p\rho)\mp \vec\tau\vec m \, J_1(p\rho)) Q(\vec p_1,\omega_1) .
\label{FFH}
\eea
Here  $\vec p=\vec p_1-\vec p_2,$ $\vec m=\vec p/ p$  and
\bea
&& J_0(y)=-8 \pi \rho^3 i_0(y),\,\, i_0(y)=\int_0^\infty dz\, \sin[zy]/zy (z \cos[\pi z/2(z^2 + 1)^{0.5}])^2,\,\,\,i(0)=0.552,
 \\
&& J_1(y)=4 \pi \rho^3 i_1(y),\,\,i_1(y)=\int_0^\infty dz\, z^2(\cos[z y]/zy - \sin[z y]/z^2y^2) \sin[\pi z/(z^2 + 1)^{0.5}].
\eea
These form-factors are localized in the region $y=p\rho\le 1 ,$ as expected. 

\section{Heavy quark light mesons interaction term}
Equation~\re{SQq} has an essential part containing the
co-product of colorless heavy-quark factor and the colorless
light-quark one:  
\bea
&&S[\psi,\psi^+,Q^+Q] =   i\lambda\int d^4x \exp(-ipx)\frac{d^3
  p_1d\omega_1}{(2\pi)^4} \frac{d^3 p_2d\omega_2}{(2\pi)^4}  
\frac{16 \pi \rho^3}{N_c}  i_0(p\rho) Q^+(\vec p_2,\omega_2) Q(\vec
p_1,\omega_1) 
\\\nonumber
&&\frac{1}{8(N_c^2-1)}\left[ \left(1-\frac{1}{2N_c}\right)\left(
q^+(x)q(x)\,\,q^+(x)q(x) + q^+\gamma_5 q\,\,q^+\gamma_5q
-q^+\vec\tau q\,\,q^+\vec\tau q-q^+\gamma_5\vec\tau
q\,\,q^+\gamma_5\vec\tau q \right)\right. 
\nonumber \\\nonumber
 &&\left.- \frac{1}{8N_c}\left(q^+\sigma_{\mu\nu}q\,\,q^+\sigma_{\mu\nu}q
  +q^+\gamma_5\sigma_{\mu\nu}q\,\,q^+\gamma_5\sigma_{\mu\nu}q 
 - q^+\sigma_{\mu\nu}\vec\tau q\,\,q^+\sigma_{\mu\nu}\vec\tau q
 - q^+\gamma_5\sigma_{\mu\nu}\vec\tau
 q\,\,q^+\gamma_5\sigma_{\mu\nu}\vec\tau q\right) \right],
\eea
where $q(x)=2\pi\rho F(i\partial )\psi(x) $ and $q^+(x)=2\pi\rho
F(i\partial )\psi^+(x). $  
 
The application of the standard bosonization procedure to the light
quarks and the calculation of the path integrals over $\lambda$ and
meson fields in the saddle point approximation lead
to the vacuum equations in the leading order (LO) on $1/N_c$
expansion.  It is natural that these mesons $(\sigma,\vec\phi, ...)$
have properties corresponding to light
quarks bilinears $(q^+q,q^+\gamma_5\vec\tau q,...)$.   

So, the vacuum equations in the LO are
written as
\bea
\frac{1}{2}\Tr \frac{iM(p)}{\hat p+i(m+M(p))}=N=\frac{1}{2}\sigma^2_0 V, \,\,\, M(p) =
MF^2(p),\,\,\, M^2= (2\pi\rho
)^4\lambda\frac{2N_c-1}{2N_c(N_c^2-1)}\sigma^2_0. 
\label{VE}\eea
They fix the coupling $\lambda$ and saddle-point $\sigma_0$ and
accordingly the dynamical quark mass $M(p)=MF^2(p)$. 

Now the total scalar meson field is $\sigma=\sigma_0+\sigma'$,
where $\sigma'$ is a quantum fluctuation. Other mesons are presented
only by their quantum fluctuations. 

At the saddle points we have the effective action for the mesons and
colorless heavy quark $Q^+Q$ bilinear as 
\bea 
\label{SQmesons}
&&S[\sigma' ,\vec\phi' ,\eta' ,\vec\sigma',Q^+Q]=
-\Tr \ln\frac{\hat p+i(m+M(p))}{\hat p+im} +N/2
\\\nonumber
&&+ \frac{1}{2}\int d^4
x\left({\sigma'}^{2}+{\vec{\phi'}}^2+{\vec{\sigma'}}^2+{\eta'}^2\right) 
-\Tr\ln\left[1+\frac{iM/\sigma_0}{\hat
    p+i(m+M(p))}F\left(\sigma'+i\gamma_5\vec\tau\vec\phi'+i\vec\tau\vec\sigma'+\gamma_5\eta'\right)F\right] 
\\\nonumber
&&-\Tr \frac{1}{\hat
  p+i(m+M(p))+\frac{iM}{\sigma_0}F\left(\sigma'+i\gamma_5\vec\tau\vec\phi'+i\vec\tau\vec\sigma'+\gamma_5\eta'\right)F} 
i\left(M(p)+\frac{M}{\sigma_0}F\left(\sigma'+i\gamma_5\vec\tau\vec\phi'+i\vec\tau\vec\sigma'+\gamma_5\eta'\right)F\right)
\\\nonumber
&&\left(\frac{i}{2} 
\int e^{-ipx}\frac{d^3 p_1d\omega_1}{(2\pi)^4} \frac{d^3 p_2d\omega_2}{(2\pi)^4} \frac{2}{N_c}J_0( p\rho)Q^+ Q\right) .
\eea
The second line describe mesons and their interactions, while the
third one explains the renormalization of the heavy
quark mass and heavy-light quark meson interactions terms. 

The renormalization of heavy quark mass  is given by 
\bea
\Delta_Q=-\Tr \frac{iM(p)}{\hat p+i(m+M(p))}
\left(\frac{i}{2} \int e^{-ipx}\frac{d^3 p_1d\omega_1}{(2\pi)^4} \frac{d^3 p_2d\omega_2}{(2\pi)^4} \frac{2}{N_c}J_0( p\rho)Q^+ Q\right) .
\eea
Taking into account Eq.~\re{VE}, we have
\bea
\Delta_Q=-\frac{2i}{N_c}\frac{1}{R^4} J_0(0)
\left( \int \frac{d^3 p_1d\omega_1}{(2\pi)^4} Q^+(\vec p_1,\omega_1)
  Q(\vec p_1,\omega_1)\right). 
\eea
So, the instanton media contribution to the heavy quark mass is
\bea
\Delta M=-2J_0(0)/N_cR^4=16\pi i_0(0)(\rho^4/R^4)\rho^{-1}/N_c,
\label{DeltaM}
\eea
in complete coincidence with~\cite{Diakonov:1989un} as we expected.

If we take the following values: $ \rho=0.35$ fm,
$R=0.856$ fm~\re{rhoR}, we obtain  
$M = 570$ MeV and $\Delta M=148$ MeV. These factors define the
coupling between heavy and light quarks in Eq.~\re{SQq}  
and certainly between heavy quarks and light mesons in Eq.~\re{SQmesons}.

\section{Conclusion}
The instanton vacuum generates a specific interaction
not only between light quarks but also between light and heavy
quarks.  All of the features of these
interaction terms are completely defined
by the instanton media parameters $\rho$ and $R$. 
It is natural to apply this approach to heavy quark and heavy-light
quark systems and predict the properties of these systems. This is a
program of our future research. 

\section{Acknowledgments}
This work was done in a close collaboration with Hyun-Chul Kim and his
group from Inha University (Korea). I am thankful to him for
many detailed discussions. 
I acknowledge the support from the grant F2-60
of Uzbekistan State Committee for Science and Technology.

\end{document}